\documentclass[twocolumn,twoside,slac_two]{revtex4}
\usepackage{graphicx}
\usepackage{fancyhdr}
\begin{document}

%Title of paper
\title{Breaking the Blazar Sequence: A new view of Radio Loud AGN
  Unification} 
\author{Eileen T. Meyer} 
\email{meyer@rice.edu}
\affiliation{Department of Physics and Astronomy, Rice University,
  Houston, TX 77005} 
\affiliation{Kavli Institute for Particle
  Astrophysics and Cosmology, 2575 Sand Hill Road, Menlo Park, CA
  94025}

\author{Giovanni Fossati}
\affiliation{Department of Physics and Astronomy, Rice University,
    Houston, TX 77005}

\author{Markos Georganopoulos}%\altaffilmark{2}}
\affiliation{Department of Physics, University of Maryland Baltimore County, Baltimore, MD 21250}
\affiliation{NASA Goddard Space Flight Center, Mail Code 663, Greenbelt, MD 20771, USA}
\author{Matthew L. Lister}%\altaffilmark{3}}
\affiliation{Department of Physics, Purdue University, West Lafayette, IN 47907}

\begin{abstract}
In recent work, we have identified two sub-populations of radio-loud
AGN which appear to be distinguished by jet structure, where
low-efficiency accreting systems produce `weak' jets which decelerate
more rapidly than the `strong' jets of black holes accreting near the
Eddington limit. The two classes are comprised of: (1) The weak jet
sources, corresponding to FR I radio galaxies, having a decelerating
or spine-sheath jet with velocity gradients, and (2) The strong jet
sources, having fast, collimated jets, and typically displaying strong
emission lines.  The dichotomy in the
$\nu_\mathrm{peak}-L_\mathrm{peak}$ plane can be understood as a
`broken power sequence' in which jets exist on one branch or the other
based on the particular accretion mode. We suggest that the intrinsic
kinetic power (as measured by low-frequency, isotropic radio
emission), the orientation, and the accretion rate of the SMBH system
are the the fundamental axes needed for unification of radio-loud AGN
by studying a well-characterized sample of several hundred
\emph{Fermi}-detected jets. Finally, we present very recent findings that the
most powerful strong jets produce gamma-rays by external Compton
rather than SSC emission, placing the dissipation region in these
strong jets at a radius inside the BLR and/or molecular torus.
\end{abstract}

%\maketitle must follow title, authors, abstract
\maketitle

\thispagestyle{fancy}

% body of paper here - Use proper section commands
% References should be done using the \cite, \ref, and \label commands
% Put \label in argument of \section for cross-referencing
%\section{\label{}}

\section{Introduction}
\label{sec:intro}

Blazars are understood to be end-on orientations of radio-loud AGN
with broad-band spectral energy distributions (SEDs) dominated by the
relativistic jet, consisting of synchrotron emission up to UV and
X-ray, and inverse Compton (IC) emission at higher energies. This
second, high-energy peak in the jet SED can be explained with both
synchrotron-self Compton (SSC, in which the particles in the
relativistic jet up-scatter synchrotron photons, \cite{mgc92,mar96})
and external Compton (EC) models, in the latter case upscattering
photons from an external source, such as the broad line region
\cite[BLR;][]{gm96,bla00,sik09}). The very high apparent luminosities,
significant polarization, and strong variability of blazars is
explained by the same Doppler boosting that makes these sources the
dominant population at high energies.

In the general unification scheme for radio-loud AGN, blazar-type
sources are beamed (\emph{i.e.}, jet face-on) counterparts to sources
seen as Fanaroff and Riley radio galaxies\citep[FR,][]{fan74}. The
association of flat-spectrum radio quasars (FSRQ) with powerful FR
IIs, and featureless BL Lac objects with FR Is was suggested chiefly
by the correspondence in spectral types, morphologies, and range in
extended radio luminosity \citep{urr95}. However, violations to this
scheme are well-known, \emph{e.g.}, powerful, FR II-like BL Lacs, and
low-power FSRQ \citep{cac04_class_testing_the_sequence,lan06,kha10},
along with BL Lacs displaying SEDs, luminous narrow lines, and
hot-spots typical of quasars \citep{ghi11_bllacs_fsrq,kha10} and even
broad lines in low continuum states \citep[\emph{e.g.}][]{ver95}. In
addition, \cite{fos98} discovered an apparently continuous
anti-correlation between the synchrotron peak (and bolometric)
luminosity ($L_\mathrm{peak}$) and the peak frequency
($\nu_\mathrm{peak}$), forming the now canonical `blazar sequence'
\cite{fos98}. With blazars displaying a continuum in
$\nu_\mathrm{peak}$ from low- to intermediate- to high- synchrotron
peaking (LSP, ISP, HSP; we use roughly $<$14, 14$-$15.5, $>$15.5 in
log $\nu_\mathrm{peak}$), it was not clear how the SED appearance of
blazars fit with the morphological dichotomy in the FR galaxies.

The blazar sequence has had several challenges. Indeed, \cite{pad03}
found that new sources they identified modify the blazar sequence to
an envelope, with the area below the blazar sequence populated with
sources. Similar envelopes were found by \cite{ant05} and
\cite{nie06}. However, lower-luminosity sources appearing in the space
below the blazar sequence were actually expected.  If one takes the
blazar sequence to be the rule for only the most-aligned (smallest
orientation angle) sources, then progressively misaligned sources will
experience less beaming and appear to form an envelope beneath the
sequence.

\begin{figure}
  \includegraphics[scale=0.5]{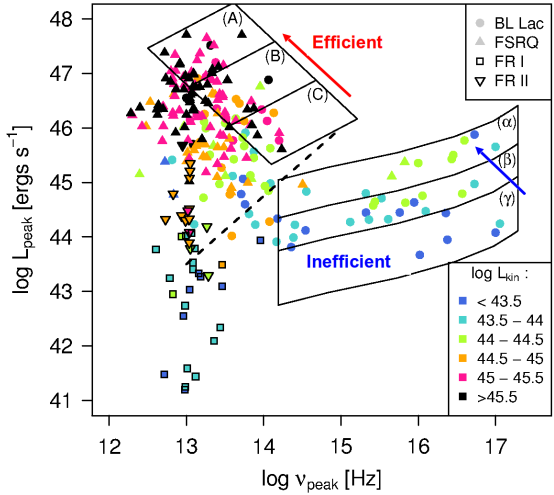}
  \caption{The synchrotron $L_{peak}$ - $\nu_{peak}$ plane. Using only
    well-sampled SEDs reveals a possible alternative to the continuous
    sequence, in which two populations show dramatically different
    behaviors in this plane. The branch at upper left is composed of
    strong-jet sources (typically powerful FSRQs), while the lower
    right branch consists of weak-jet sources (typically low-power BL
    Lacs). The upper contour of the box around the weak-jet sources is
    typical of a decelerating jet model characterized by velocity
    gradients. The two sets of boxes are used to bin sources for
    further analysis, as discussed in Section
    \ref{sec:explore}. \emph{Adapted from M11.}}
  \label{mainenv}
\end{figure}

The possibility for tracking similar jets as they filled the envelope
motivated recent work \cite[][hereafter M11]{mey11}, in which we
filled the synchrotron $L_\mathrm{peak}-\nu_\mathrm{peak}$ plane using
a much larger sample of well-characterized jet SEDs from sources over
a wide range of orientations (from blazars to radio galaxy jets imaged
with high-resolution instruments). We tested whether the jet kinetic
power ($L_\mathrm{kin}$, as measured from the isotropic radio
emission) could be the single parameter necessary for characterizing
the jet SED, along with radio core dominance ($R$ = log
$L_\mathrm{core}$/$L_{ext}$) as a measure of jet orientation (and
therefore degree of beaming). \emph{We found that the blazar sequence
  is actually broken into two populations in the $L_\mathrm{peak} -
  \nu_\mathrm{peak}$ plane (Fig.~\ref{mainenv}).} The `weak' jets
(branch labeled `inefficient') consist entirely of sources with
$L_\mathrm{kin}<$10$^{44.5}$ ergs s$^{-1}$, and extend from
mis-aligned FR I radio galaxies at low $\nu_\mathrm{peak}$ up to
well-aligned (as measured by $R$) HSP BL Lacs. The `strong' jets
comprised a population of low $\nu_\mathrm{peak}$ jets (branch labeled
`efficient') which drop in $L_\mathrm{peak}$ rapidly with decreasing
$R$.

This manuscript is organized as follows. In Section \ref{sec:mdot} we
discuss the three (minimum) parameters that we believe necessary for
radio-loud AGN unification, including a critical transition in
accretion efficiency. In Section \ref{sec:explore} we show that the
blazar divide of M11 is consistent with a \emph{broken} power sequence
with two branches. In Section \ref{sec:blls} we briefly discuss the
recurring issue of FSRQs masquerading as BL Lacs, and the potential
pitfalls of dividing blazars by optical type only. Finally, in Section
\ref{sec:fermi} we introduce the `Blazar Envelope' as seen by
\emph{Fermi} and discuss the importance of the jet kinetic power and
orientation of the jet in determining the gamma-ray luminosity.  We
also examine a recent finding suggesting that the high-energy emission
of high-power jets is dominated by EC processes rather than SSC.

\section{A Critical Accretion Rate?}
\label{sec:mdot}

Following M11, the important factors in RL AGN unification (\emph{i.e.},
placing a source in Fig.~\ref{mainenv}) appear to be (a) the jet power
$L_\mathrm{kin}$ (above 10$^{44.5}$ erg s$^{-1}$, one only finds
sources on the strong-jet branch) and (b) the orientation angle (as
proxied by radio core dominance in this case). However, there is a
degeneracy at low $L_\mathrm{kin}$, with sources appearing in both
branches, which would imply that a third parameter is necessary
(optical type is not reliable, as discussed in
Section~\ref{sec:blls}). Others have suggested that a critical
accretion rate \citep{ghi01,ghi09}, taking place at $\dot{m}$ =
$\dot{m}_\mathrm{cr}\sim$10$^{-3}-$10$^{−2}m_\mathrm{Edd}$
\citep{nar97,nar08}, and marking a transition from sources with
inefficiently radiating accretion disks and therefore absent or weak
broad emission lines (\emph{i.e.}, BLL/FR Is) to sources with efficient
accretion disks and strong broad emission lines (FSRQ/FR IIs), could
explain the two distinct classes (optical/morphological).

\begin{figure}
  \includegraphics[scale=0.5]{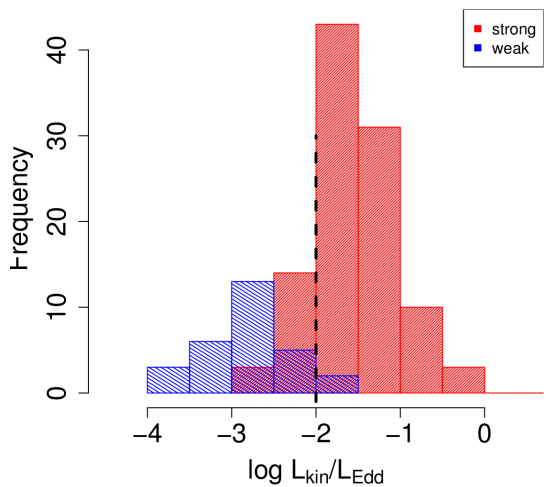}
  \caption{\small Histogram of estimated accretion rate
    ($L_\mathrm{kin}$/$L_\mathrm{Edd}$) for a sample of blazars with
    known jet power and a reliable estimate of black hole mass.  The
    sample is sub-divided into weak-jet sources (from
    Fig.~\ref{mainenv}, those with log $\nu_\mathrm{peak}>$14.5),
    and strong-jet sources (those with log $\nu_\mathrm{peak}<$14.5
    and log $L_\mathrm{peak}>$45.5).  Strong-jet sources have log
    $L_\mathrm{kin}$/$L_\mathrm{Edd}$ typically greater than critical
    values of -2 to -3 suggested in the literature, and thus this
    branch is labeled 'efficient' in Fig.~\ref{mainenv}; while
    weak-jet sources have sub-critical values and are labeled
    `inefficient'.}
  \label{lkinledd}
\end{figure}

In Fig.~\ref{lkinledd}, we compare the estimated accretion rates for
two sub-samples of the sources analyzed in M11 (\emph{i.e.}, those in
Fig.~\ref{mainenv}). The accretion rate $\dot{m}$ is estimated from the
ratio $L_\mathrm{kin}$/$L_\mathrm{Edd}$, where $L_\mathrm{kin}$ is the
jet power scaled from low-frequency isotropic radio emission (M11) and
$L_\mathrm{Edd}$ is the Eddington luminosity $L_\mathrm{Edd}$ =
1.3x10$^{38}$\,($M$/$M_\odot$).  Black hole masses $M$ were taken from
the literature and averaged (further details will be discussed in a
forthcoming paper). The two populations were carefully selected to
correspond to strong-jet and weak-jet samples while minimizing
contamination. To that end, the weak-jet sample was chosen from all
sources with known synchrotron log $\nu_\mathrm{peak}$ greater than
14.5, and the strong-jet sample was chosen from all sources with log
$\nu_\mathrm{peak}<$14.5 \emph{and} log $L_\mathrm{peak}>$45.5.
The latter cut was necessary to avoid potentially confusing LSP
sources shown in the lower left of Fig.~\ref{mainenv}. Because sources
on either branch, when misaligned, begin to coincide in this space,
these sources were ambiguous in classification and omitted from the
analysis.

As shown in Fig.~\ref{lkinledd}, there does indeed appear to be a
divide at $-$log $L_\mathrm{kin}$/$L_\mathrm{Edd}$ $\sim$ 2$-$3.  The weak-jet sources appear at
lower accretion rates and the strong-jet sources at higher. Despite
the fact that the errors on black hole masses and jet powers can be
over half an order of magnitude, the divide in the populations is
fairly distinct. Some have suggested that the accretion power merely
gives an upper bound to $L_\mathrm{kin}$ \citep{fer11}; this would
then predict that the strong jet sources should have a significant
population appearing at sub-critical efficiency as estimated for
Fig.~\ref{lkinledd}. We suggest that a one-to-one correspondence is
more in keeping with the results shown there, and that the accretion
rate as measured by $L_\mathrm{kin}$/$L_\mathrm{Edd}$ can be used as
the third factor in AGN unification.

\section{Exploring the new Broken Sequence}
\label{sec:explore}

In M11, the new scheme was presented as a broken blazar sequence, with
two broad groups, otherwise undifferentiated.  We return to the
synchrotron $\nu_\mathrm{peak}-L_\mathrm{peak}$ plane to examine the
possibility that a `blazar sequence' remains within each of the two
branches (\emph{i.e.}, a strong sequence and a weak sequence).

\begin{figure}
  \includegraphics[width=3.4in]{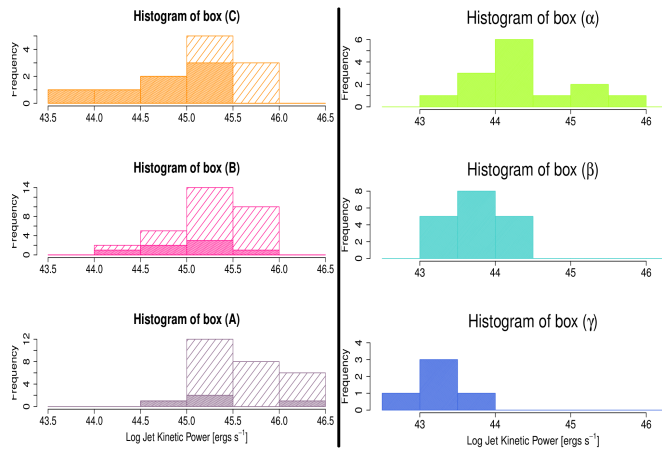}
  \caption{\small (Left) Distributions of $L_\mathrm{kin}$ for the
    boxes labeled A,B,C in Fig.~\ref{mainenv}.  BL Lacs are shown as
    shaded regions. (Right) Distributions of $L_\mathrm{kin}$ for the
    boxes labeled $\alpha$,$\beta$,$\gamma$ in Fig.~\ref{mainenv}. In
    both cases the predicted change in $L_\mathrm{kin}$ along a
    sequence is seen, supporting a `broken' power sequence
    interpretation.}
  \label{combohist}
\end{figure}

\noindent
\textbf{Building a Sequence.} For simplicity, consider all sources to
have a black hole mass $M\sim$10$^9\,M_\odot$, and thus a fixed
$L_\mathrm{Edd}$. Sources that accrete at $\dot{m} < \dot{m}_{cr}$
will have radiatively inefficient accretion disks. Assuming that the
jet power cannot significantly exceed the accretion power, we require
$L_\mathrm{kin}\le\dot{m}L_\mathrm{Edd} <
\dot{m}_{cr}L_\mathrm{Edd}$. Sources that accrete at $\dot{m} >
\dot{m}_{cr}$will have radiatively efficient accretion disks and
$L_\mathrm{kin} < \dot{m}L_\mathrm{Edd}$. In both cases, given a set
of sources with identical $L_\mathrm{kin}$, the aligned (0$^\circ$
orientation) sources will have the highest $L_\mathrm{peak}$ and
$\nu_\mathrm{peak}$.  Because $\nu_\mathrm{peak}$ decreases as the
radiative cooling becomes stronger, any situation where electrons in
more powerful sources experience stronger radiative cooling will
result to a \emph{decreasing $\nu_\mathrm{peak}$ with increasing
  $L_\mathrm{kin}$}, \emph{i.e.}, a sequence.  If we now gradually increase
$L_\mathrm{kin}$/$L_\mathrm{Edd}$, the location of
$L_\mathrm{peak}-\nu_\mathrm{peak}$ for our 0$^\circ$ aligned source
on the plot shifts to delineate a power sequence. The exact shape of
this track depends on the physics of the jet.  We consider below two
plausible cases for the weak-jet and strong-jet populations
separately.

\noindent
\textbf{The Inefficient Accretion Case.} 
We consider here
the case in which the jet bulk Lorentz factor ($\Gamma$) does
not change as $L_{kin}$  is increased, and $L_{kin}$ scales linearly
with the number density of relativistic electrons in the jet.
In this case, and assuming that the electron energy distribution is
$n(\gamma)=k \gamma^{-p}, \; p\le 3$,  $\nu_{peak}$  is formed by the
maximum energy electrons and remains fixed, while the peak luminosity
increases linearly with the number density of relativistic electrons and,
therefore, with  $L_{kin}$.
%
%We consider here the case in
%which the jet bulk Lorentz factor ($\Gamma$) does not change as
%$L_\mathrm{kin}$ is increased, and $L_\mathrm{kin}$ scales linearly
%with the number density of relativistic electrons and the magnetic
%field energy density. If n($\gamma$) = k$\gamma^{-2}$ is the injected
%electron distribution with the electron Lorentz factor ($\gamma$) and
%magnetic field $B$, $k\propto B^{1/2}\propto L_\mathrm{kin}$. A break
%at $\gamma_b$ will form in the steady-state electron distribution
%where the cooling and the escape times are equal, and $\gamma_b
%\propto B^{-2}$, where we assumed that synchrotron cooling, and
%therefore synchrotron luminosity, is comparable to or stronger than
%SSC cooling, and therefore SSC luminosity. Emission at
%$L_\mathrm{peak}$ is produced by electrons at $\gamma_b$ and
%$L_\mathrm{peak}\propto
%k\gamma_b^{-2}B^{-2}\gamma_b^{2}\delta^3\propto kB^2 \propto
%L_\mathrm{kin}^2$, where we used our assumption that $\Gamma$ does not
%change and the jet is aligned at 0$^\circ$. We then have
%$\nu_\mathrm{peak}\propto B\gamma_b^2\delta\propto B B^{-4}\propto
%B^{-3}\propto L_\mathrm{kin}^{-3/2}$ . We therefore have for the
%inefficient accretion jets the following sequence:
%$L_\mathrm{peak}\propto\nu_\mathrm{peak}^{-4/3}\propto
%L_\mathrm{kin}^2$. 
%
This is the line forming the upper right of the box for the
inefficient branch in Fig.~\ref{mainenv}.  Similar lines, possibly of
different slope, are anticipated for other scenarios in which cooling
increases with increasing $L_\mathrm{kin}$. However, an important
feature is that when $L_\mathrm{kin}$ increases enough that the source
passes $\dot{m}_{cr}$, the inefficient accretion mode ceases being
attainable. Interestingly, the highest $L_\mathrm{kin}$ sources in the
inefficient accretion branch reach values of 10$^{44.5}$ erg s$^{-1}$,
which corresponds to $\dot{m}_{cr}\sim2.3\times10^{-3}$ for $M$ =
10$^9M_\odot$.

Does $L_\mathrm{kin}$ actually increase along the inefficient
branch sequence?  To evaluate this, we assume that the de-beaming
tracks shown in Fig.~\ref{mainenv} (curved lines defining the box at
lower right) are a good representation of the actual de-beaming taking
place. Keeping only HSP and ISP, we plot the distribution of $L_{kin}$
in the three zones $\alpha$,$\beta$,$\gamma$ in Fig.~\ref{combohist}
(right panel). As expected, the average $L_\mathrm{kin}$ decreases as
we move `down' the sequence.

\noindent
\textbf{The Efficient Accretion Case.} A plausible scenario for the
strong-jet sources is that $L_\mathrm{kin}$ increases with $\Gamma$,
as suggested by VLBI data \citep[\emph{e.g.}][]{kha10}. We assume in addition
that the comoving energy density of electrons and magnetic field
remain invariant. In this case, $L_\mathrm{kin}\propto
\Gamma^2$. Because the external photon field energy density in the
comoving frame scales as $U\propto U_\mathrm{ext}$ and $U_\mathrm{ext}$
is not a function of $\dot{m}$, the cooling break in the electron
distribution scales as $\gamma_b \propto 1/U \propto
\Gamma^{-2}$. Then $\nu_\mathrm{peak} \propto B\gamma_b^2\delta
\propto \Gamma^{-3}$, where we have used our assumption that $B$ is
fixed and our jet is aligned. Now $L_\mathrm{peak} \propto \Gamma^4$
for emission from plasma moving through a steady feature in the
jet. Therefore, we have the sequence $L_\mathrm{peak} \propto
\nu_\mathrm{peak}^{-4/3}\propto L_\mathrm{kin}^2$, shown as the red
arrow in Fig.~\ref{mainenv}. 

To evaluate if $L_\mathrm{kin}$ indeed increases along the path shown,
we considered the sources in boxes A, B, and C. The distribution of
$L_\mathrm{kin}$ is shown for each box in the left panel of
Fig.~\ref{combohist}. As can be seen, the average $L_\mathrm{kin}$
increases from C to B to A, as expected.

\section{A Caution on `False BL Lacs'}
\label{sec:blls}
As noted by \citep{geo98} and most recently
\citep{ghi11_bllacs_fsrq,gio11}, there is a possibility that some
broad-lined blazars appear as lineless sources (and therefore, are
classified falsely as BL Lacs), due to the blazar jet being much
brighter than any lines in the optical.  This seems to require a
combination of a source being (a) somewhat more beamed (so as to
enhance the jet emission relative to the isotropic line emission) and
(b) with a synchrotron peak which is at least near to the optical
(\emph{i.e.}, possibly, a lower-power source). In keeping with the
idea of a half-sequence still existing in the strong-jet population,
one would expect that more powerful sources have lower
$\nu_\mathrm{peak}$ values, and brighter broad line spectra.
Following aligned jets in the powerful branch from low to high
$L_\mathrm{kin}$, $\nu_\mathrm{peak}$ will shift to lower energies and
out of the optical, and the lines would come `up', the source would be
revealed as an FSRQ rather than BL Lac.  The net effect, which we can
readily test, should be that in moving `up' the strong-jet sequence
(red arrow in Fig.~\ref{mainenv}), the fraction of sources recorded as
BL Lacs will drop.  In fact this is seen in the left panel of
Fig.~\ref{combohist}, where we show BL Lac sources as the dark-shaded
regions of the histogram.  It is clear that in the progression from
box C to B to A that the percentage of BL Lacs drops (58\% to 24\% to
15\%), as expected if the broad lines of these sources are simply
overcome by a bright jet in optical.

\begin{figure}
  \includegraphics[width=3in]{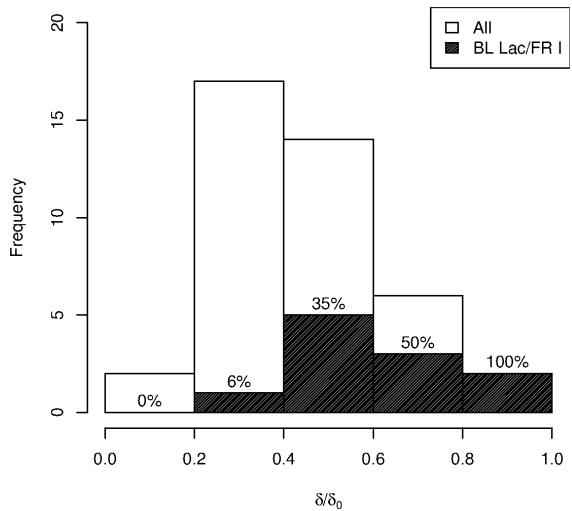}
  \caption{\small For moderate power (10$^{43.5}-$10$^{44.5}$ erg
    s$^{-1}$) sources in the strong-jet branch, we show the
    distribution of the relative beaming factor $\delta$/$\delta_0$ as
    estimated from the position of the source relative to the sequence
    derived in Section \ref{sec:explore}. The shaded regions correspond to
    BL Lac sources.  The fraction of BL Lacs falls off as beaming
    decreases, suggesting that these are false BL Lacs due to the jet
    emission covering the lines.}
  \label{blls}
\end{figure}

Further, we can check the effect of relative beaming on the proportion
of BL Lacs in the strong-jet branch. Taking the scaling derived in
Section~\ref{sec:explore}, we can draw a line (with somewhat arbitrary
normalization) corresponding to the sequence at 0 degree alignment
(\emph{e.g.}, the red line in Fig.~\ref{mainenv}). Selecting sources at
moderate power (10$^{43.5}-$10$^{44.5}$ erg s$^{-1}$) to get a more
uniform population (see M11), we can estimate the relative beaming by
considering that a departure from the 0$^\circ$ line will follow a
de-beaming path in the log-log plot with a slope of 4, \emph{i.e.}, the
relative beaming factor $\delta$/$\delta_0$ can be estimated from the
ratio $\nu$/$\nu_0$, where $\nu_0$ is the frequency where the sequence
and de-beaming path meet.  As a source is found further from the red
line, the smaller the value of $\delta$/$\delta_0$. As shown in
Fig.~\ref{blls}, the fraction of BL Lacs falls from 100\% in the
highest bin of $\delta$/$\delta_0$ down to 0\%, which seems to support
the idea that \emph{beaming} is also an important factor in producing
false BL Lacs.

\section{Fermi and the Blazar Divide}
\label{sec:fermi}

\subsection{The Blazar Envelope at High Energies}
\label{ssec:benv}

\begin{figure}
  \includegraphics[width=3in]{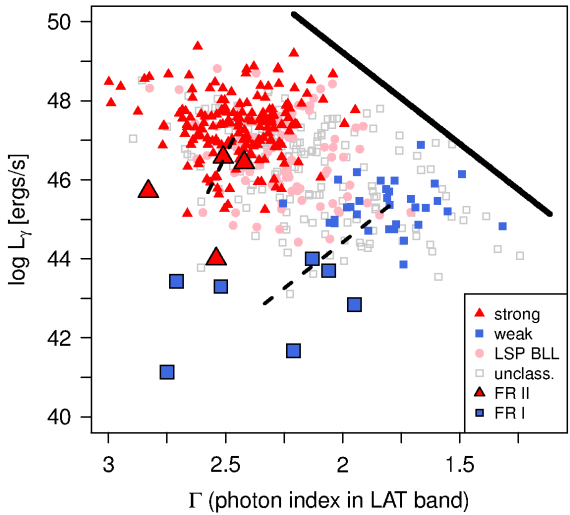}
  \caption{\small The Inverse Compton Envelope.  Figure is adapted
    from \cite{abd10_misaligned}.  The \emph{Fermi}-detected AGN
    (large symbols) show that alignment plays a strong effect in the
    gamma-ray output.  Interestingly, while the FR II sources appear
    to drop directly below the powerful FSRQ sources, FR I radio
    galaxies appear to follow a more horizontal track, similar to our
    findings in the synchrotron envelope (Fig.~\ref{mainenv}). The
    presence of a “forbidden zone” (empty region at upper right,
    implying no powerful, hard sources) suggests that there is a
    sequence in the IC plane, similar to that found in the
    synchrotron.}
  \label{icenv}
\end{figure}

Given that the trends shown in Fig.~\ref{mainenv} appear to be driven
by jet power and beaming (\emph{i.e.}, orientation angle), it is natural to
expect something similar to appear in the \emph{inverse Compton}
$L_{peak}$ - $\nu_{peak}$ plane. We show in Fig.~\ref{icenv} the total
LAT-band Luminosity (k-corrected) versus the LAT-band photon index
($\Gamma$) for blazars in the second \emph{Fermi} catalog
\citep[2FGL;][]{abd11_2fgl}. While each of these measurements are very
rough approximations for what we wish to measure (IC
$\mathrm{L}_\mathrm{peak}$ and $\nu_\mathrm{peak}$, respectively), we
still see a familiar trend: at low $\Gamma$ values (corresponding to
higher IC $\nu_\mathrm{peak}$), the gamma-ray luminosities are lower,
suggesting again a zone of avoidance at high-peak, high-luminosity.
The presence of two branches can not be readily discerned, but this
may be in part due to the roughness of the estimators and
particularly, the choice of the LAT-band integrated luminosity as a
proxy for $\mathrm{L}_\mathrm{peak}$, as this will progressively
underestimate sources peaking beyond the LAT band on either side
(before for powerful FSRQ, and after for the HSP sources). It may also
be due to the fact that the IC spectrum is not necessarily produced
the same way for all sources (\emph{i.e.}, the Compton dominance over
synchrotron may vary).

An important feature in this figure is the location of the radio
galaxies.  As discussed above, we suggest that weak-jet sources are to
be paired with FR I radio galaxies.  As seen in Fig.~\ref{icenv}, the
FR I sources are found to have substantially softer LAT spectra than
the ISP/HSP sources (marked with blue squares).  This seems to match
the horizontal trend discussed for Fig.~\ref{mainenv}, which can be
explained if the jets have velocity gradients. On the other hand, the
FR II radio galaxies seen with \emph{Fermi} appear to have very
similar IC peak frequencies to the strong-jet sources, consistent with
a `simple' jet as discussed in M11.  In both cases a dotted line is
shown connecting the average $L_{peak}$, $\nu_{peak}$ for the blazars
and their respective parent population of radio galaxies.

\begin{figure}
  \includegraphics[width=3in]{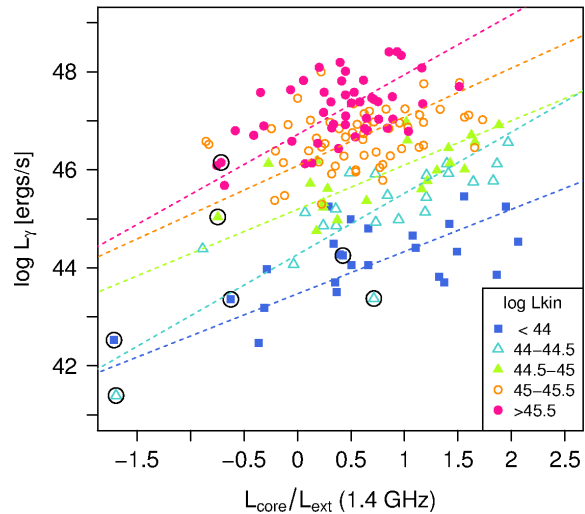}
  \caption{\small Total LAT-band gamma ray luminosity versus radio
    core dominance for the 2FGL sources (including radio galaxies at
    lower left) with good SED coverage.  Jet power (shown by
    color/symbol changes) has a very strong effect on the gamma-ray
    luminosity, but when sources are binned by $L_\mathrm{kin}$, the
    dependence on radio core dominance also becomes apparent. The
    lines shown are OLS bisector fits, all of which show significant
    positive correlation.}
  \label{lgvr}
\end{figure}

In Fig.~\ref{lgvr}, we give a different view of the high-energy
emission of blazars as seen with \emph{Fermi}. Using a sub-sample of
the 2FGL blazars with very good spectral coverage (in order to
accurately fit a phenomenological model to determine the
synchrotron/IC peaks), we show the average IC peak luminosity versus
the radio core dominance at 1.4 GHz ($R$). For the whole sample, there
is no trend; however, when the sample is divided into bins of
$L_\mathrm{kin}$, the correlation between the peak $L_\gamma$ and $R$
(log $L_\mathrm{core}$/$L_\mathrm{ext}$) become significant within the
bands.  We have shown the OLS bisector fits for each bin as dashed
lines (slopes range from 0.8 - 1.2, with positive Pearson's r values
from 0.4 - 0.8). The peak IC luminosity is strongly positively
correlated with both the jet power and $R$, as we expect if these two
factors are key to predicting jet phenomenology.

\subsection{Diagnosing SSC versus EC with Populations}
\label{ssec:sscec}

The high-energy emission in blazars can be explained with either SSC
or EC emission. Generally low-power and lineless objects (\emph{i.e.}, weak
jets) are believed to radiate by SSC at high energies, while
individually, many high-powered FSRQ (strong jets) have been more
satisfactorily fit with EC models \cite{ghi10,ver11}.  However a
consistent basis for which sources require EC has not been
demonstrated for any particular class of blazars.

The location of the GeV emission is a matter of active debate, as it
comes from locations close to the central engine that remain
unresolved. However, identifying the nature of the radiation can give
some insight. If the emission region is located within the sub-pc
scale BLR, the GeV emission of blazars with strong lines should have a
strong EC component from up-scattered BLR photons
\cite{sik94}. However, if the region is several pc downstream (well
outside the BLR/Molecular Torus) as suggested by some multiwavelength
observations of variability \citep{bot09,agu11}, EC is necessarily
ruled out.

\begin{figure}
  \includegraphics[width=3in]{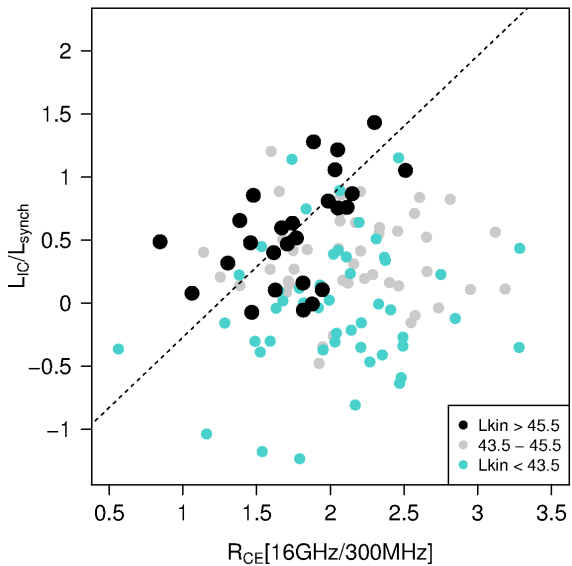}
  \caption{\small Inverse Compton dominance over synchrotron emission
    ($L_\mathrm{IC}$/$L_\mathrm{sync}$) versus the radio core
    dominance $R$. The sources shown are 2FGL blazars which were
    monitored by the Owens Valley Telescope at 16 GHz, in order to get
    contemporaneous measurements.  We find that only for the strongest
    jets (black), the inverse Compton dominance increases with
    increasing alignment, as measured by $R$, consistent with EC
    emission processes.  The other sources show no trends.}
  \label{lgls}
\end{figure}

In Figure \ref{lgls} we examine the relationship between Compton
Dominance ($CD$, the log-ratio of IC and synchrotron
$L_\mathrm{peak}$) and the radio core dominance, $R$, for a sample of
2FGL blazars.  The black points are high-jet-power sources (log
$\mathrm{L}_\mathrm{kin}$ $>$ 45.5), while the blue are low-power (log
$\mathrm{L}_\mathrm{kin}$ $<$ 43.5).  The flat distribution in the
latter group may be consistent with the SSC models generally used to model
these types of sources. For the high-power group, however, we find a
significant trend of increasing Compton dominance with increasing
radio core dominance (\emph{i.e.}, alignment). \emph{This can be interpreted
  as a signature of an EC process for a population of jets
  characterized by high kinetic powers}.

This statement can be explained as follows. The Doppler beaming
factor, $\delta$, is a function of the Lorentz factor $\Gamma$ and the
orientation angle, and for monochromatic luminosities we have
$\mathrm{L}$ = $\mathrm{L}_0\delta^{3+\alpha}$, where $L_0$ is the
rest-frame luminosity (at $\delta = 1$), and $\alpha$ is the energy
spectral index at the frequency of interest.  The exponent 3+$\alpha$
is the value assumed for a `moving blob' in the jet. If the emission
comes from a standing shock, the exponent would be 2+$\alpha$
\cite{bla85}.

In the case of emission by SSC, the IC peak has a beaming pattern
which is identical to the synchrotron (\emph{i.e.}, it follows the above
equation). However, for EC models, the beaming at high energies goes
as $ \mathrm{L}_\mathrm{IC}$ =
$\mathrm{L}_\mathrm{0,IC}\delta^{4+2\alpha}$ \cite{der95,geo01}. The
larger exponent indicates that as a source with significant EC
emission is aligned, the IC peak should be more and more dominant over
the synchrotron peak (\emph{i.e.}, $CD$ will increase with $R$. We can see
that $CD \propto $log ($\delta^{(1+\alpha)}$), and thus
the dependence on $\delta$ goes with an exponent of 2 when comparing
the IC and synchrotron peaks (where $\alpha$ = 1). At radio
frequencies, where the spectral index is confined to a fairly narrow
range in values of $\alpha_r = 0 - 0.5$, the beaming exponent will be
$\sim$ 3 - 3.5. Thus an overall slope of $\sim$ 2/3, up to 1 is
expected in Fig.~\ref{lgls}.  While a more detailed investigation is
in progress, we note that a higher slope might mean that the
assumption that the Lorentz factor of the plasma emitting in the
gamma-rays is the same as that emitting at GHz frequencies is
incorrect. While the current data shown are insufficient to put strong
constraints on the slope, there is sufficient confidence in the
positive correlation to say that EC is required for the most powerful
strong-jet sources.

%\section{Summary}
%%\label{sec:sum}%
%
%
%We have shown that the \emph{Fermi} gamma-ray satellite sees
%relativistic jets in RL AGN over a wide range of orientation angles,
%including sources seen as radio galaxies.  There is a `forbidden zone'
%of high-power, high-peak sources which remains empty, consistent with
%a spectral sequence similar to that seen in the synchrotron peak
%luminosity - peak frequency plane. Further, the much lower IC peak
%frequencies found for the few detected FR I radio galaxies suggest
%that the low-power class of FR I/BL Lacs debeam along more horizontal
%tracks in the synchrotron and IC planes (Figures 1 and 2). We show
%that the total gamma-ray band luminosity depends on both orientation
%(measured through radio core dominance) and more heavily on the jet
%kinetic power (as measured from low-frequency, isotropic radio
%emission). Finally, we present the first collective evidence for EC
%process in a sample of high-power jets (log $\mathrm{L}_\mathrm{kin}$
%$>$ 45.5), showing that as alignment (radio core dominance) increases,
%the Compton dominance increases dramatically.

% If you have acknowledgments, this puts in the proper section head.
\bigskip % extra skip inserted
\begin{acknowledgments}
GF and EM acknowledge support from NASA grants NNG05GJ10G, NNX06AE92G,
and NNX09AR04G, as well as SAO grants GO3-4147X and G05-6115X. MG
acknowledges support from the NASA ATFP grant NNX08AG77G and NASA
FERMI grant NNH08ZDA001N. The MOJAVE project is supported under
National Science Foundation grant 0807860-AST.
\end{acknowledgments}

\bigskip % extra skip inserted
% Create the reference section using BibTeX:
%\bibliography{basename of .bib file}
%\begin{thebibliography}{9}   % Use for  1-9  references

\end{document}